# Channel Estimation for MIMO MC-CDMA Systems

K.Sureshkumar[1], R.Rajalakshmi[1], A.Vetrikanimozhi[2]

AP/ECE CK College of Engineering and Technology[1], Lecturer/ECE, IFET College of Engineering[2]

`m.k.sureshkoumar@gmail.com`[1], `raji.ckcet@ymail.com`[1], `vetri0909@gmail.com`[2]

*Abstract*

*The concepts of MIMO MC-CDMA are not new but the new technologies to improve their functioning are an emerging area of research. In general, most mobile communication systems transmit bits of information in the radio space to the receiver. The radio channels in mobile radio systems are usually multipath fading channels, which cause inter-symbol interference (ISI) in the received signal. To remove ISI from the signal, there is a need of strong equalizer. In this thesis we have focused on simulating the MIMO MC-CDMA systems in MATLAB and designed the channel estimation for them.*

## 1. Introduction

Code division multiple access (CDMA) is a multiple access technique where different users share the same physical medium, that is, the same frequency band, at the same time. The main ingredient of CDMA is the *spread spectrum* technique, which uses high rate signature pulses to enhance the signal bandwidth far beyond what is necessary for a given data rate. In a CDMA system, the different users can be identified and, hopefully, separated at the receiver by means of their characteristic individual *signature pulses* (sometimes called the *signature waveforms*), that is, by their individual *codes*. Nowadays, the most prominent applications of CDMA are mobile communication systems like CDMA One (IS-95), UMTS or CDMA 2000. To apply CDMA in a mobile radio environment, specific additional methods are required to be implemented in all these systems [4]. Methods such as power control and soft handover have to be applied to control the interference by other users and to be able to separate the users by their respective codes.

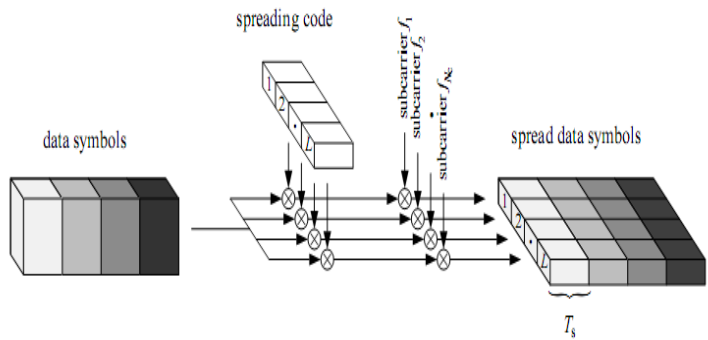

Fig 1. *Signal generation of MC-CDMA*

In a spread spectrum technique to enhance the signal bandwidth far beyond what is necessary for a given data rate and thereby reducing the power spectral density (PSD) of the useful signal so that it may even sink below the noise level. One can imagine that this is a desirable property for military communications because it helps to hide the signal and it makes the signal more robust against intended interference (*jamming*). Spreading is achieved – loosely speaking – by a multiplication of the data symbols by a *spreading sequence* of pseudo random signs. These sequences are called *pseudo noise* (PN) sequences or code signals. In CDMA networks there are a number of channels each of which supports a very large number of users. For each channel the base station generates a unique code that changes for every user. The base station adds together all the coded transmissions for every subscriber. The subscriber unit correctly generates its own matching code and uses it to extract the appropriate signals [4].

In order for all this to occur, the pseudo-random code must have the following properties: It must be deterministic. The subscriber station must be able to independently generate the code that matches the base station code. It must appear random to a listener without prior knowledge of the code (i.e. it has the statistical properties of sampled white noise). The cross-correlation between any two codes must be small. The code must have a long period (i.e. a long time before the code repeats itself). Generation of PN sequence: The p-n sequence is usually generated using a shift register with feedback-taps. Binary sequences are shifted through the shift registers in response to clock pulses and the output of the various stages are logically combined and fed back as input to the stage. When the feedback logic consists of exclusive –OR gates, the shift register is called a linear PN sequence generator. The initial contents of the memory stages and the feedback logic circuit determine the successive contents of the memory. If a linear shift register reaches zero state at some time, it would always remain in the zero state and the output would subsequently be all zeros. Since there are exactly 2m-1 nonzero state for an m-stage feedback shift register, the period of a PN sequence produced by a linear m-stage shift register cannot exceed 2m-1 symbol which is called maximal length sequence.

## 2. Multiple Inputs and Multiple outputs

The Recent works have shown that multiple input multiple output (MIMO) systems can achieve an increased capacity without the need of increasing the operational bandwidth. Also for the fixed transmission rate, they are able to improve the signal transmission quality (Bit Error Rate) by using spatial diversity. In order to obtain these advantages MIMO systems require accurate channel state information (CSI) at least at the receiver side [1]. This information is in the form of complex channel matrix.

The method employing training sequences is a popular and efficient channel estimation method. A number of training based channel estimation methods for MIMO systems have been proposed, as shown in [2], [3]. However, in most of the presented works independent identically distributed (i.i.d) Rayleigh channels are assumed [2] [3]. This assumption is rarely fulfilled in practice, as spatial channel correlation occurs in most of propagation environments. In an MMSE channel estimator for MIMO MC-CDMA was developed and its performance was tested under spatial correlated channel. However, a very simple correlated channel model was used. These investigations neglected the issue of antenna array used at the receiver side.

In practical cases there is a demand for small spacing of array antenna elements at least at the mobile side of MIMO system. This is required to make the transceiver of compact size. However, the resulting tight spacing is responsible for channel correlation. Also, the received signals are affected by mutual coupling effects of the array elements. This paper investigates the case of a MIMO wireless system, in which the signal is transmitted from a fixed transmitter to a mobile terminal receiver equipped with a uniform circular array (UCA) antenna.

The investigations make use of the assumption that the signals arriving at this array have an angle of arrival (AoA) that follows a Laplacian distribution. Also taken into account are mutual coupling effects in UCA. For this system, we test the performance of MMSE channel estimation method, which was described in [2] [3]. The investigations concern such factors as antenna spacing, and decay factor which is related to angle spread.

## 3. System Model & MMSE Channel Estimator

### A. SYSTEM MODEL

A flat block-fading narrow-band MIMO system with Mt transmit antennas and Mr receive antennas is considered. Later on, Mr value is fixed to 4. The relation between the received signals and the training sequences is given by (1):

$$Y = HP + V \qquad (1)$$

where Y is the Mt x N complex matrix representing the received signals, P is the Mt x N complex training matrix,
which includes training sequences (pilot signals); H is the Mr x Mt complex channel matrix and V is the Mr x N complex zeromean white noise matrix.

### B. MMSE ESTIMATION

The goal is to estimate the complex matrix H from the knowledge of Y and P. Assuming the training matrix is known, the channel matrix can be estimated using the minimum meansquare error (MMSE) method, as described in [2] [3]

$$\bar{H} = \frac{\rho}{M_t} Y P^H (R_H^{-1} + \frac{\rho}{M_t} P P^H)^{-1} \qquad (2)$$

with MSE estimation error given by:

$$J_{MMSE} = E\{ \|H - H_{MMSE}\|_F^2 \} = tr\left\{ \left( R_H^{-1} + \frac{\rho}{M_t} P P^H \right)^{-1} \right\} \qquad (3)$$

Where p is the signal to noise ratio, E {.} is a statistical expectation and tr {.} denotes the trace of matrix, $\| \|_F^2$ stands for the Frobenius norm and the channel correlation matrix is given by $R_H = E\{H^H H\}$

## 4. Channel Estimation

Based on those assumptions such as perfect synchronization and block fading, we end up with a compact and simple signal model for both the single antenna MC CDMA and MIMO-MC CDMA systems. In training based channel estimation algorithms, training symbols or pilot tones that are known to the receiver, are multiplexed along with the data stream for channel estimation. The idea behind these methods is to exploit knowledge of transmitted pilot symbols at the receiver to estimate the channel.

For a block fading channel, where the channel is constant over a few MC CDMA symbols, the pilots are transmitted on all subcarriers in periodic intervals of MC CDMA blocks. This type of pilot arrangement, depicted in Fig. (2), is called the block type arrangement. For a fast fading channel, where the channel changes between adjacent MC CDMA symbols, the pilots are transmitted at all times but with an even spacing on the subcarriers, representing a comb type pilot placement, Fig. (3) The channel estimates from the pilot subcarriers are interpolated to estimate the channel at the data subcarriers.

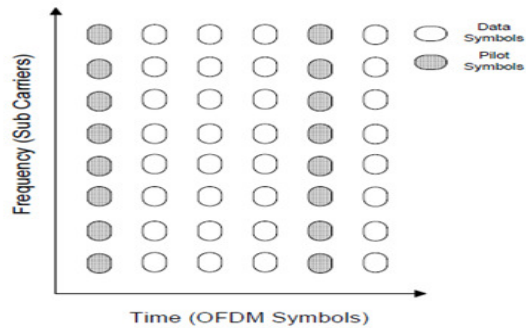

Fig: 2 Block type pilot arrangements

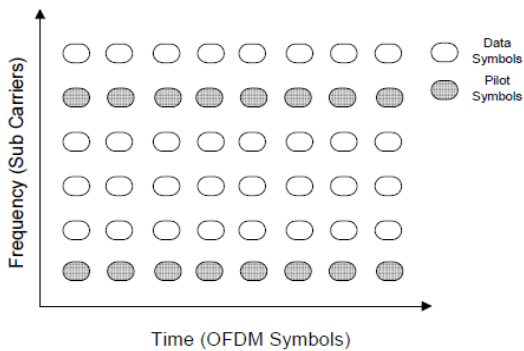

Fig: 3 Comb type pilot arrangements

In block-type pilot based channel estimation, OFDM channel estimation symbols are transmitted periodically, in which all subcarriers are used as pilots. If the channel is constant during the block, there will be no channel estimation error since the pilots are sent at all carriers. The estimation can be performed by using either LS or MMSE [9], [10]. In comb-type pilot based channel estimation, the $Np$ pilot signals are uniformly inserted into $X(k)$ according to the following equation:

$$X(k) = X(mL + l) \quad l=1,2,\ldots L-1$$

Where $L$ = No. of subcarriers / $Np$ and $m$ is pilot carrier index. If inter symbol interference is eliminated by the guard interval, we write (6) in matrix notation $Y = XFh + W$

### A. SYSTEM ENVIRONMENT

Wireless The system environment we will be considering in this thesis will be wireless indoor and urban areas, where the path between transmitter and receiver is blocked by various objects and obstacles. For example, an indoor environment has walls and furniture, while the outdoor environment contains buildings and trees. This can be characterized by the impulse response in a wireless environment.

### B. MULTIPATH FADING

Most indoor and urban areas do not have direct line of sight propagation between the transmitter and receiver. Multi-path occurs as a result of reflections and diffractions by objects of the transmitted signal in a wireless environment. These objects can be such things as buildings and trees. The reflected signals arrive with random phase offsets as each reflection follows a different path to the receiver. The signal power of the waves also decreases as the distance increases. The result is random signal fading as these reflections destructively and constructively superimpose on each other. The degree of fading will depend on the delay spread (or phase offset) and their relative signal power.

### C. FADING EFFECTS DUE TO MULTI-PATH FADING

Time dispersion due to multi-path leads to either flat fading or frequency selective fading: Flat fading occurs when the delay is less than the symbol period and affects all frequencies equally. This type of fading changes the gain of the signal but not the spectrum. This is known as amplitude varying channels or narrowband channels. signal is narrow compared to the channel bandwidth. Frequency selective fading occurs when the delay is larger than the symbol period. In the frequency domain, certain frequencies will have greater gain than others frequencies.

## 5. Numerical Result

In this section, we evaluate the processing gain to estimate the channel conditions. Figure 4 shows comparison of MMSE with LMSE. Here we consider MIMO based system for high multimedia communication system. The parameter consider for simulation is (number of transmitter) nt=2 and (number of receiver) nr=2 and (processing gain) pg=32. Fig 4 expounds the MSE for MMSE and LS estimation. It can be seen from the plot that MMSE results is better performance compared to LMSE estimation.

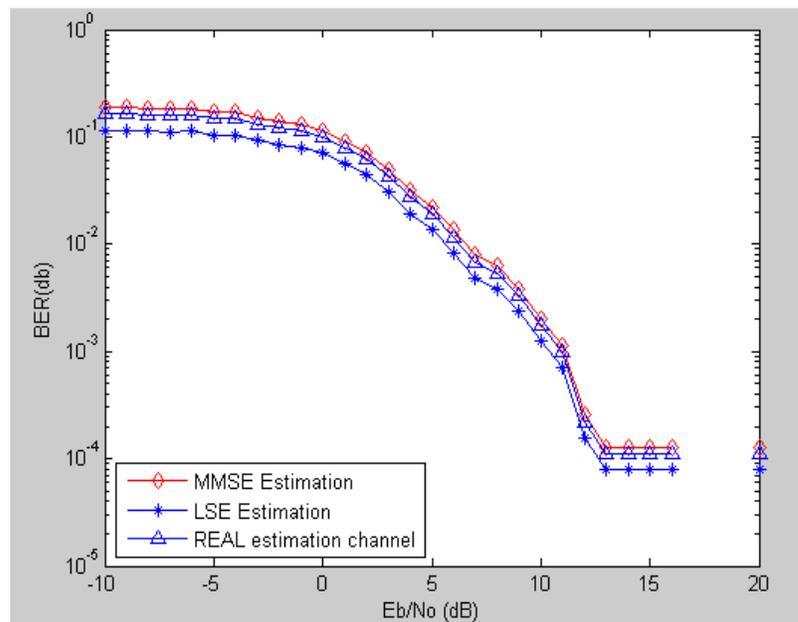

Fig 4. *BER and SNR plots for MC CDMA*

In this section, we evaluate the processing gain to estimate the channel conditions. Figure 5 shows comparison of MMSE with LMSE. Here we consider MIMO based system for high multimedia communication system.

The parameter consider for simulation is (number of transmitter) nt=2 and (number of receiver) nr=2 and (processing gain) pg=16. Fig 5 expounds the MSE for MMSE and LS estimation. It can be seen from the plot that MMSE results is better performance compared to LMSE estimation.

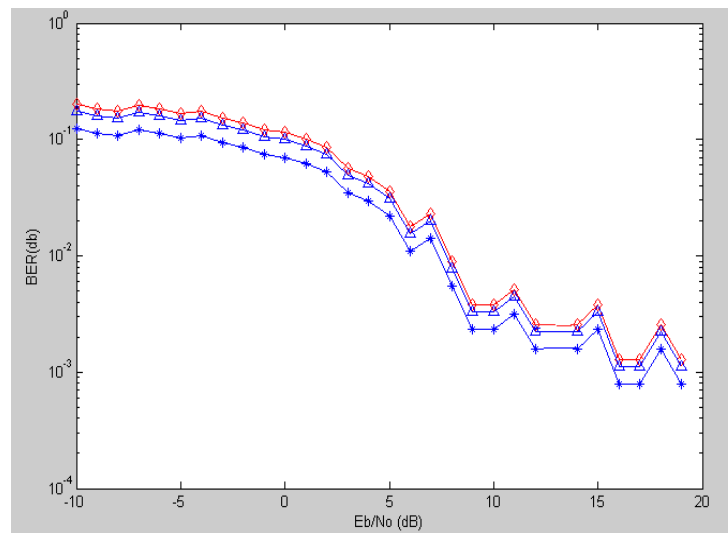

Fig 5. *BER and SNR plots for MC CDMA*

In this section, we evaluate the processing gain to estimate the channel conditions. Figure 6 shows comparison of MMSE with LMSE. Here we consider MIMO based system for high multimedia communication system. The parameter consider for simulation is (number of

352

transmitter) nt=2 and (number of receiver) nr=3 and (processing gain) pg=32. Fig 6 expounds the MSE for MMSE and LS estimation. It can be seen from the plot that MMSE results is better performance compared to LMSE estimation

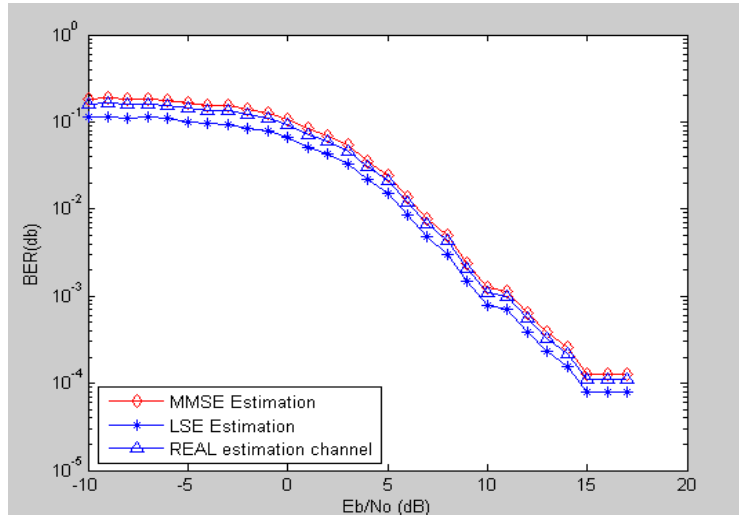

Fig 6 *BER and SNR plots for MC CDMA*

Figure 7 shows comparison of MMSE with LMSE . Here we consider MIMO based system for high multimedia communication system. The parameter consider for simulation is (number of transmitter) nt=2 and (number of receiver) nr=4 and (processing gain) pg=32. Fig 7 expounds the MSE for MMSE and LS estimation. It can be seen from the plot that MMSE results is better performance compared to LMSE estimation.

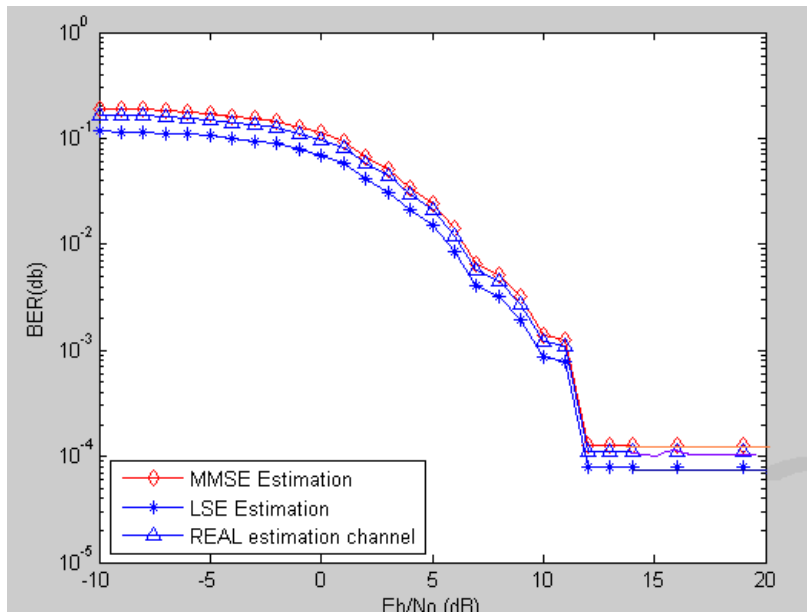

Fig 7 *BER and SNR plots for MC CDMA*

Figure.8 shows the MSE plot for MMSE and LS estimation techniques. The MMSE estimation technique outperforms LS technique.

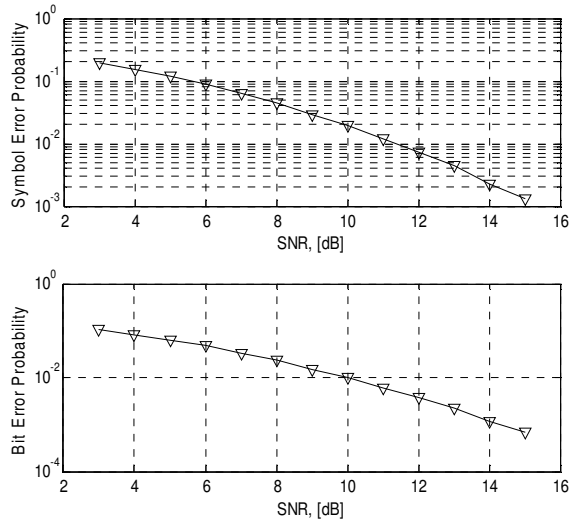

Fig *8 BER and SER plots for MC CDMA using MMSE estimators*

The estimators use on pilot data that is known to both transmitter and receiver as a reference in order to track the fading channel. The estimators use block based pilot symbols, meaning that pilot symbols are sent across all sub-carriers periodically during channel estimation. This estimate is then valid for one MIMO/MC-CDMA frame before a new channel estimate will be required. we simulated MC- CDMA systems, the BER Vs SNR plots for the previously mentioned systems has obtained and satisfactory is shown in Figure 8. Then the simulation of channel estimation is carried out using MMSE estimators and the ,BER and SER plots were obtained.

In this paper channel estimation based on both block-type pilot in MIMO MC-CDMA based systems are compared. Channel estimation based on block-type pilot arrangement is achieved by giving the channel estimation methods at the pilot frequencies and the interpolation of the channel at data frequencies. The estimators in this study can be used to efficiently estimate the channel in both MC-CDMA systems given certain knowledge about the channel statistics. The MMSE estimator assumes a *priori* knowledge of noise variance and channel covariance. We then discussed Space Time Block coding and maximum likelihood decoding in the MIMO MC-CDMA system to enhance its performance further. We can also observe the advantage of diversity in MIMO system results less BER than SISO system. And simulation results show that MMSE estimation for MIMO MC-CDMA provides less MSE than other systems. Finally, by comparing the performance of MMSE with LS, it is observed that the former is more resistant to the noise in terms of the channel estimation.

## 6. Conclusion

Thus we have focused on simulating the MIMO and MC-CDMA systems in MATLAB and designed the channel estimation. The MIMO is used to exploit the features of MC-CDMA such as, maximum spectral efficiency, easy implementation and simplicity along with spatial diversity. We first design pilot structure for basic channel estimation. We then discuss enhanced channel estimation by catching significant taps and discarding noisy taps, which significantly improves the performance of channel estimation. To obtain channel parameters corresponding to data blocks, we develop optimal interpolation approach to substitute simple linear interpolation. We have analyzed the impact of channel estimation error and found that the performance of and MC-CDMA can be further improved by more accurate channel estimation.